\documentclass[journal, letter, 9pt]{IEEEtran}
\usepackage{amsmath}
\usepackage{amsfonts}
\usepackage{graphics}
\usepackage{array}
\usepackage{shortvrb}
\usepackage{epsf}
\usepackage{graphicx}
\usepackage{rotating}
\usepackage{multirow}
\usepackage{cite}
\usepackage{color}
\usepackage{alltt}
\usepackage{soul}
\usepackage{overpic}
\usepackage{pict2e}
\usepackage{comment}
\usepackage{acronym}
\usepackage{float}
\usepackage[caption=false,font=footnotesize]{subfig}
\usepackage{placeins}
\usepackage{threeparttable}
\usepackage{booktabs}

\acrodef{cig}[CIG]{converter-interfaced generator}
\acrodef{coi}[COI]{Center of Inertia}
\acrodef{avr}[AVR]{Automatic Voltage Regulator}
\acrodef{agc}[AGC]{Automatic Generation Control}
\acrodef{lic}[LIC]{Local Integration Control}
\acrodef{pfc}[PFC]{Primary Frequency Control}
\acrodef{sm}[SM]{synchronous machine}
\acrodef{der}[DER]{Distributed Energy Resource}
\acrodef{ffr}[FFR]{Fast Frequency Regulation}
\acrodef{pll}[PLL]{Phased-Locked Loop}
\acrodef{rocof}[RoCoF]{Rate of Change of Frequency}
\acrodef{dae}[DAE]{Differential-Algebraic Equation}
\acrodef{pi}[PI]{Proportional-Integral}
\acrodef{pod}[POD]{Power Oscillation Damper}
\acrodef{pf}[PF]{Participation Factor}
\acrodef{lep}[LEP]{Linear Eigenvalue Problem}
\acrodef{ess}[ESS]{Energy Storage system}
\begin{document}
\title{Enhancing Frequency Control through Rate of Change of Voltage Feedback}
\newcommand{\T}{^{\scriptscriptstyle \rm T}}
\newcommand{\wcoi}{\omega_{\scriptscriptstyle \rm COI}}
\newcommand{\dwcoi}{\dot\omega_{\scriptscriptstyle \rm COI}}
\newcommand{\vd}{v_{d}}
\newcommand{\vq}{v_{q}}
\newcommand{\dvd}{\dot{v}_{d}}
\newcommand{\dvq}{\dot{v}_{q}}
\newcommand{\vdt}{v'_{d}}
\newcommand{\vqt}{v'_{q}}
\newcommand{\dvdt}{\dot{v}'_{d}}
\newcommand{\dvqt}{\dot{v}'_{q}}
\newcommand{\wt}{\tilde{\omega}}

\newcommand{\jj}{\jmath}

\author{
  Federico Milano, {\em IEEE Fellow}, Bibi Alhanjari, Georgios
  Tzounas, {\em IEEE Member} %
  \thanks{F.~Milano, B.~Alhanjari and G.~Tzounas are with the School
    of Electrical \& Electronic Engineering, University College
    Dublin, D04V1W8, Dublin, Ireland. (e-mails:
    \mbox{federico.milano@ucd.ie}, %
    \mbox{bibi.ghjmalhanjari@ucdconnect.ie}, %
    \mbox{georgios.tzounas@ucd.ie}).}%
  \vspace{-7mm} }

\markboth{Submitted to IEEE Transactions on Power Systems}{}
\maketitle
\begin{abstract}
  This letter proposes a simple and inexpensive technique to improve
  the frequency control of distributed energy resources.  The proposed
  control consists in modifying the conventional estimated bus
  frequency signal with an additional feedback signal that utilizes
  the rate of change of the voltage magnitude measured at the same
  bus.  The case study showcases the benefits of the proposed control
  and compares its performance with standard frequency control schemes
  through time-domain simulations.
\end{abstract}
\IEEEpeerreviewmaketitle

\section{Introduction}
\label{sec:intro}

The increasing proliferation of converter-based generation is known to
reduce the available mechanical inertia in the electric power grid.
As a result, frequency variations triggered from imbalances between
power supply and demand become more prominent and faster, and threaten
the system's stability and performance.  In response to this
challenge, system operators and researchers have been actively seeking
for techniques that enhance the effectiveness of frequency control
services provided by \acp{der}.

Literature in the field has explored several optimal control and
coordination strategies for different converter-based energy
technologies, including wind, solar PV, storage, and demand response
systems \cite{bevrani2021power, obaid2019frequency}.  For example,
recent works have proposed analytical methods for the design of
synthetic inertia and droop coefficients of \acp{der} to meet a
desired dynamic performance, as well as optimal \ac{der} placement and
power sharing strategies, e.g.~see \cite{dhople:2018a,
  poolla2019placement}.

In previous work, the authors of this paper have studied the
analytical links between power, frequency and voltage variations in
transmission and distribution networks to establish alternative
control signals that can improve the stability and primary response of
low-inertia systems \cite{DER,sanniti2022curvature}.  In this vein,
this letter proposes to modify the bus frequency signal conventionally
utilized in \ac{der} primary control loops to include as additional
feedback signal the rate of change of the voltage magnitude measured
at the same bus.
The rationale of the proposed control scheme is based on a recent work
by the first author that defines a novel quantity, namely, the
\textit{complex frequency}, where the rate of change of the voltage
magnitude constitutes the real part and the conventional frequency is
the imaginary part \cite{freqcomplex}.

% ======================================================================
\section{Rationale and Proposed Control Scheme} 
\label{sec:theory}
% ======================================================================

The definition of the complex frequency relies on a general property
of complex quantities, as follows.  Let us consider the Park vector of
the bus voltage, say $\bar{v}$, as time-dependent complex value that
utilizes the $dq$-axis components of the Park reference frame rotating
at constant angular speed $\omega_o$,~i.e:
\begin{equation}
  \label{v:park}
  \bar{v}(t) = v_d(t) + j \, v_q(t) \, ,
\end{equation}
where $j$ is the imaginary unit.  This voltage is substantially a
dynamic phasor and can be written in polar coordinates as:
\begin{equation}
  \label{eq:v1}
  \bar{v} = v \, e^{j \, \theta} \, ,
\end{equation}
where we have dropped the dependency on time for economy in notation.
Defining $u = \ln(v)$, $v \ne 0$, \eqref{eq:v1} becomes:
\begin{equation}
  \label{eq:v2}
  \bar{v} = e^{u + j \, \theta} \, .
\end{equation}
If $\bar{v}$ is a function of time, then the derivative of
\eqref{eq:v2} leads to:
\begin{equation}
  \label{eq:vdot}
  \dot{\bar{v}}
  = (\dot{u} + j \, \dot{\theta}) \, e^{u + j \, \theta}
  = (\dot{u} + j \, \dot{\theta}) \, \bar{v} \, .
\end{equation}
Equaling \eqref{eq:v1} and \eqref{v:park} and taking into account the
rotation of the Park reference frame, one has:
\begin{align}
  \label{eq:omega}
  \omega &= \dot{\theta} =
           \frac{\vd \dvq - \vq \dvd}{v^2} + \omega_o \, , \\
  \label{eq:rho}
  \rho &= \dot{u} = \frac{\dot{v}}{v} =
         \frac{\vd \dvd + \vq \dvq}{v^2} \, ,
\end{align}
where $\omega$ is the conventional instantaneous frequency of
$\bar{v}$ and $\rho$ can be defined as an \textit{instantaneous
  bandwidth} \cite{Cohen:1995} or, using a geometric analogy, a
\textit{radial frequency} \cite{frenet}.
The time derivative of the Park vector in \eqref{v:park} can be
written as:
\begin{equation}
  \label{eq:eta}
  \dot{\bar{v}} =
  (\rho + j \, \omega) \, \bar{v} =
  \bar{\eta} \, \bar{v}\, ,
\end{equation}
where $\bar{\eta}$ is the \textit{complex frequency} as defined in
\cite{freqcomplex}.

For the discussion of the control proposed in this letter, it is
relevant to rewrite $\omega$ and $\rho$ in \eqref{eq:omega} and
\eqref{eq:rho} assuming now that the Park transform is obtained using
the frequency of the \ac{coi}, say $\wcoi$, rather than the
synchronous reference frame $\omega_o$.  This leads to:
\begin{align}
  \label{eq:omega2}
  \omega &= \frac{\vdt \dvqt - \vqt \dvdt}{v^2} + \wcoi \, , \\
  \label{eq:rho2}
  \rho &= \frac{\vdt \dvdt + \vqt \dvqt}{v^2} \, ,
\end{align}
where $\vdt + j \vqt$ is the Park vector obtained for the $dq$-axis
reference frame rotating at $\wcoi$.

While $\vdt \ne \vd$ and $\vqt \ne \vq$, the values of $\omega$ and
$\rho$ on the left-hand sides of \eqref{eq:omega2} and \eqref{eq:rho2}
are in effect equal to the values in \eqref{eq:omega} and
\eqref{eq:rho}, respectively, as $\omega$ and $\rho$ are geometric
invariants, that is, their values are independent from the coordinates
utilized to measure the components of the voltage.  The invariance of
$\rho$ is straightforward to show, as the identity
$v = |\vd + j \vq| = |\vdt + j \vqt|$ must hold independently of the
reference speed chosen to obtain the Park transform.  Demonstrating
the invariance of $\omega$ is more involved, and the interested reader
can find a proof in \cite{geom}.

The invariance of $\omega$ and $\rho$ leads to the following remarks:
\begin{itemize}
\item Equation \eqref{eq:omega2} shows that the instantaneous
  frequency $\omega$ can be decomposed into two terms.  The first
  captures exclusively local bus dynamics and is null in steady state.
  The second, i.e. $\wcoi$, is slow and follows the system-wide
  dynamic of the frequency.
\item The radial frequency $\rho$ depends only on local dynamics and,
  hence, is always null in steady state.
\end{itemize}

It is immediate to observe that, if the voltage at the bus where the
frequency is measured is regulated through an automatic control, then
the local variations of $\rho$ can be relatively small w.r.t. the
variations of $\omega$.  An obvious limit case is that of an ideal
voltage controller, for which $\rho = 0$ as the voltage is always kept
perfectly constant.  This also justifies some models of the bus
voltage signal proposed in the literature \cite{Karpilow:2022}.
However, since ideal voltage controllers do not exist in practice, we
can always assume that the voltage magnitude and hence also $\rho$ do
exhibit a transient behavior following a disturbance.

In this letter we show the benefits for the frequency control of power
systems of including $\rho$ in the conventional frequency control
input signal.  Such control is meaningful when the dynamics of
$\omega$ and $\rho$ evolve in similar time scales.  This is
particularly relevant for converter-interfaced devices, whose
frequency controllers can be designed to be as fast or faster than
their voltage regulation.  We exploit this feature to design the
proposed \ac{der} control.  On the other hand, the proposed control is
not suitable for conventional synchronous machines, as their voltage
regulators are much faster than the dynamics that can be tracked by
turbine governors.

A consequence of assuming similar time scales of frequency and voltage
controllers is that the local terms of $\omega$ and $\rho$ in
\eqref{eq:omega2} and \eqref{eq:rho2}, while having different
expressions, have similar harmonic content and thus, show a similar
``trend''.  This statement is further illustrated in the case study
presented in Section \ref{sec:cstudy} but we also justify it
qualitatively below with an analytic example.  Let us assume that:
\begin{equation}
  \label{eq:ex}
  \begin{aligned}
    \vdt &= V - k \, e^{-\alpha t} \cos(\beta t) \, , \\
    \vqt &= k \, e^{-\alpha t} \sin(\beta t) \, ,
  \end{aligned}
\end{equation}
where $V$, $k$, $\alpha$ and $\beta$ are constant.  The expressions in
\eqref{eq:ex} resemble a typical frequency transient in power systems
if $k \ll V$ and $\alpha, \beta \ll \wcoi$, where $\alpha$ and $\beta$
represent the damping and the angular frequency, respectively, of the
dominant mode of the system.  Leveraging these inequalities, one can
get to the following approximated expressions from \eqref{eq:omega2}
and \eqref{eq:rho2}:
\begin{equation}
  \label{eq:ex2}
  \begin{aligned}
    \omega - \wcoi &\approx \frac{k \, e^{-\alpha t}}{V} \,
    [\beta \cos(\beta t) - \alpha \sin(\beta t)] \, , \\
    \rho &\approx \frac{k \, e^{-\alpha t}}{V} \,
    [\beta \sin(\beta t) + \alpha \cos(\beta t)] \, .
  \end{aligned}
\end{equation}
The latter expressions show that $\rho$ and $\omega - \wcoi$ have
same order of magnitude and show same oscillatory behavior.

We use the observations and the empirical result above as follows.
Conventional frequency controllers of non-synchronous resources
utilize the estimation of the frequency of the voltage at their point
of connection with the grid, e.g., using a \ac{pll}, and compare it
with a reference, typically $\omega_o$.  The resulting frequency error
signal is thus a mix of the local frequency oscillations and the
system-wide frequency deviation due to the power imbalance in the
grid.  This means that the local and system-wide variations are
weighted in the same way by the controller.  This appears inevitable
since, in order to estimate the local term of the instantaneous
frequency in \eqref{eq:omega2}, one would need to be able to measure
$\wcoi$ first, which is though not available to local controllers.
However, extrapolating the result of the example above and based on
the experience matured on a large number of simulations, we have
observed that we can decouple, even if in an approximated way, these
two effects and ``weight'' them differently in the frequency control.

The main advantage of the rate of change of voltage ($\rho$) as
compensating signal for the frequency control is that it is a purely
\textit{local} signal, as opposed to the frequency that is both local
and system-wide as it intrinsically contains information on the
frequency of the center of inertia ($\wcoi$).  Controllers that utlize
the rate of change of frequency (RoCoF) are effective as long as the
variations of the RoCoF are not biased by the variations of $\wcoi$.
In conventional systems or systems where converters emulate the
behavior and time scales of synchronous machines, e.g.,
\cite{uros2018}, the variations of $\wcoi$ are slower than local
frequency oscillations, and that is why controllers based on the RoCoF
can be effective.  However, in systems with very low inertia, $\wcoi$
can partially overlap with local fluctuations and thus lead to a less
effective control based on the RoCoF.  We note, moreover, that
controllers based on the RoCoF generally require an additional control
channel in parallel with the conventional frequency droop control.  On
the other hand, $\rho$ can be included in any existing \ac{der}
frequency controller.

In summary, we propose to use $\rho$ to build the following modified
input signal to primary frequency controllers:
\begin{equation}
  \label{eq:wt}
  \wt = \omega - K \, \rho \, ,
\end{equation}
where $K$ is the parameter to be adjusted and that allows tuning the
impact of local frequency oscillations on the power output of the
converter-interfaced device.  It is relevant to observe that the
estimation of $\rho$ is simple and readily available as, from
\eqref{eq:rho}, one simply needs to measure $v$ and estimate
$\dot{v}$.

% ======================================================================
\section{Case Study}
\label{sec:cstudy}
% ======================================================================

In this section, we illustrate the effect of the proposed control with
the WSCC 9-bus system and the New England 39-bus system.  All
simulation results are obtained with the software tool Dome
\cite{dome}.  In the simulations, estimations of $\omega$ and $\rho$
are obtained using a synchronous-reference frame \ac{pll} and voltage
measurements at the point of connection of the \ac{cig}.

\subsection{WSCC 9-bus System}

\begin{figure}[ht!]
  \begin{center}
    \resizebox{0.80\linewidth}{!}{\includegraphics[scale=1.0]{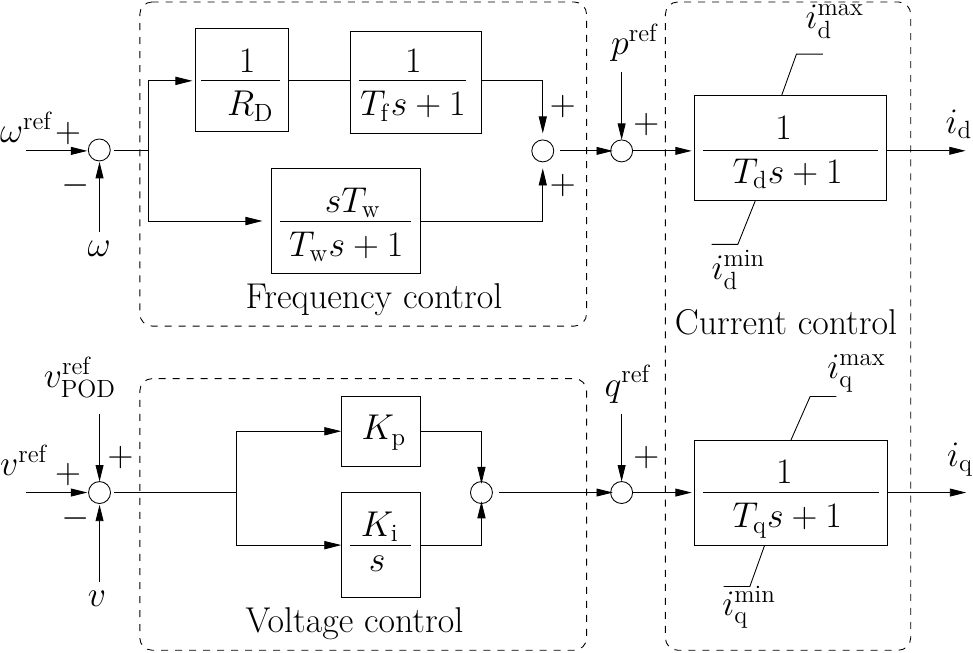}}
    \caption{Control diagram of the primary controllers of \acp{der}.
      The conventional frequency controller utilizes the signal
      $\omega$, whereas the proposed control utilizes as input signal
      $\tilde{\omega}$ as defined in \eqref{eq:wt}.}
    \label{fig:der}
  \end{center}
  \vspace{-3mm}
\end{figure}

The original WSCC 9-bus system is modified to emulate a low-inertia
system by reducing the inertia constants of the \acp{sm}, namely, 4~s
for \ac{sm} 1 and 2 and 3~s for \ac{sm} 3.
Then, a \ac{cig} is connected through a transformer to bus~7.  The
block diagram of the \ac{cig} control is shown in Fig.~\ref{fig:der}.
The inner loop regulates the $dq$-axis currents and includes limiters.
The outer loop consists of a voltage controller and a frequency
controller with droop and washout channels.  In the scheme,
$p^{\rm ref}$ and $q^{\rm ref}$ are the operating set points for a
given period of the \ac{der} as defined by the market and/or
transmission system operators; whereas
$v^{\rm ref}_{\scriptscriptstyle \rm POD}$ is an auxiliary signal
coming from power oscillator damper (POD) connected to the DER, which
is equivalent ot the power system stabilizers of conventional
synchronous generators.  The considered \ac{der} utilizes a
grid-following converter.  It is important to note that employing
grid-forming converters would not change the effectiveness of the
proposed compensating signal based on $\rho$ as long as the voltage
and frequency controllers have similar time scales.  At the initial
operating point the \ac{cig} generates 100~MW, which are accommodated
in the system by reducing by the same amount the power produced by
\ac{sm}~2.

To study the effectiveness of the proposed control in improving the
frequency response of the system, we first compare the observability
that the signals $\omega$ and $\tilde \omega$ provide to the dynamic
modes of the system directly linked to primary frequency regulation.
Recall that a dynamic mode represents a primary frequency control mode
if it meets the following properties \cite{moeini2016analytical}: (i)
it is global, i.e.~all buses are coherent to the mode.  (ii) the
associated mode shapes for all generator speeds are in phase; and
(iii) its natural frequency lies in the range 0.02-0.1~Hz.

\begin{figure}[ht!]
  \centering
  \resizebox{0.49\linewidth}{!}{\includegraphics[scale=1.0]{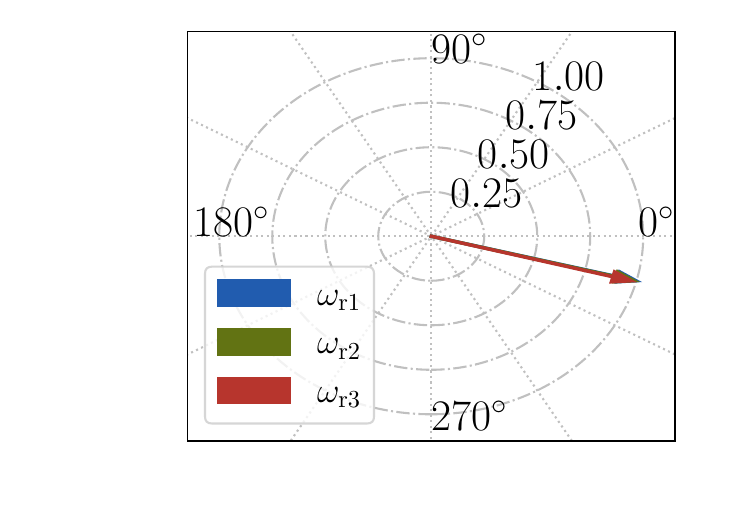}}
  \resizebox{0.48\linewidth}{!}{\includegraphics[scale=1.0]{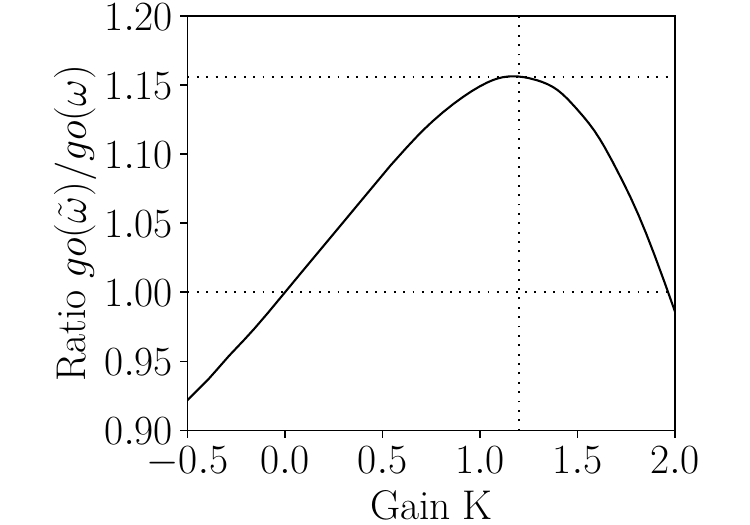}}
  \vspace{-3mm}
  \caption{WSCC test system: Left: Frequency control mode shapes of
    \ac{sm} speeds.  Right: geometric observability ratio for the
    conventional and proposed frequency control modes as a function of
    $K$. }
  \label{fig:track} 
\end{figure}

We carry out a small-signal analysis of the system assuming that the
\ac{cig} frequency control loop is switched off.  The results indicate
that the system is stable at the examined equilibrium and that the
primary frequency control mode is represented by the complex pair of
eigenvalues $-0.55\pm \jj 1.12$ with natural frequency 0.09~Hz.  The
corresponding normalized mode shapes of the rotor speeds of the three
\acp{sm} are depicted in the left panel of Fig.~\ref{fig:track}.

The geometric observability $go$ \cite{eigbook:2021} of the frequency
control mode by $\rho$, $\omega$, and $\tilde \omega$ for $K=1$, are
summarized in Table~\ref{tab:obs}.  The right panel in
Fig.~\ref{fig:track} tracks the ratio $go(\tilde \omega)/go(\omega)$
between the observability of $\tilde \omega$ and $\omega$, as a
function of $K$.  This analysis confirms that $\rho$ is effective to
improve the dynamic behavior of the system.  Note, however, that the
specific value of the gain depends on the estimation of $\rho$, which
in the simulations below is obtained through a PLL that estimates
$\dot{v}$, and the parameters of the frequency controller of the CIG.

\begin{table}[!t]
  \centering
  \footnotesize 
  \renewcommand{\arraystretch}{1.05}
  \caption{Geometric observability $go$ of frequency control mode.}
  \label{tab:obs}
  \begin{tabular}{l|ccccc}
    \toprule
    Signal 
    & $\rho$ & $\omega$ & $\tilde \omega=\omega - \rho$
    \\
    \midrule
    Observability & 0.34 & 0.87 & 1.00
    \\ \bottomrule
  \end{tabular}
\end{table}

\begin{figure}[ht!]
  \centering
  \resizebox{0.49\linewidth}{!}{\includegraphics[scale=1.0]{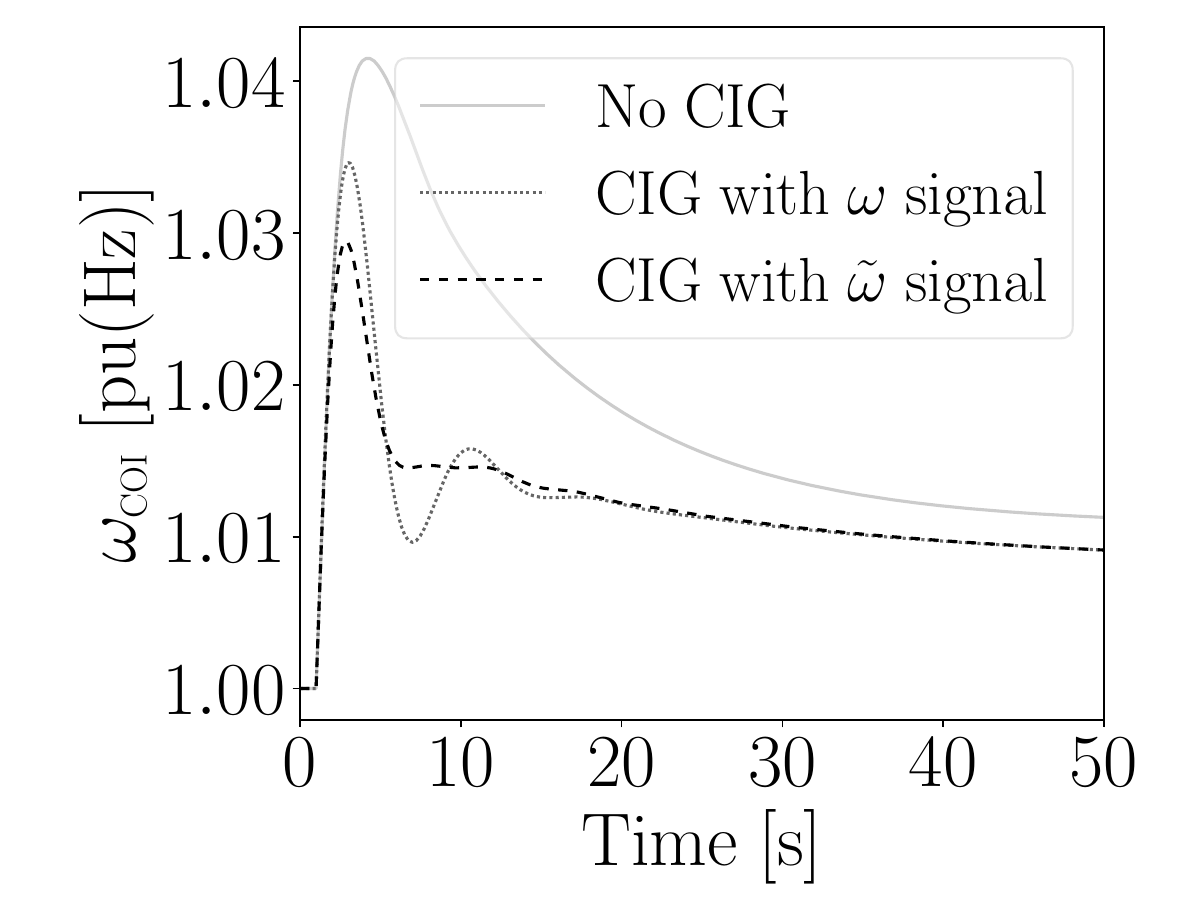}}
  \resizebox{0.49\linewidth}{!}{\includegraphics[scale=1.0]{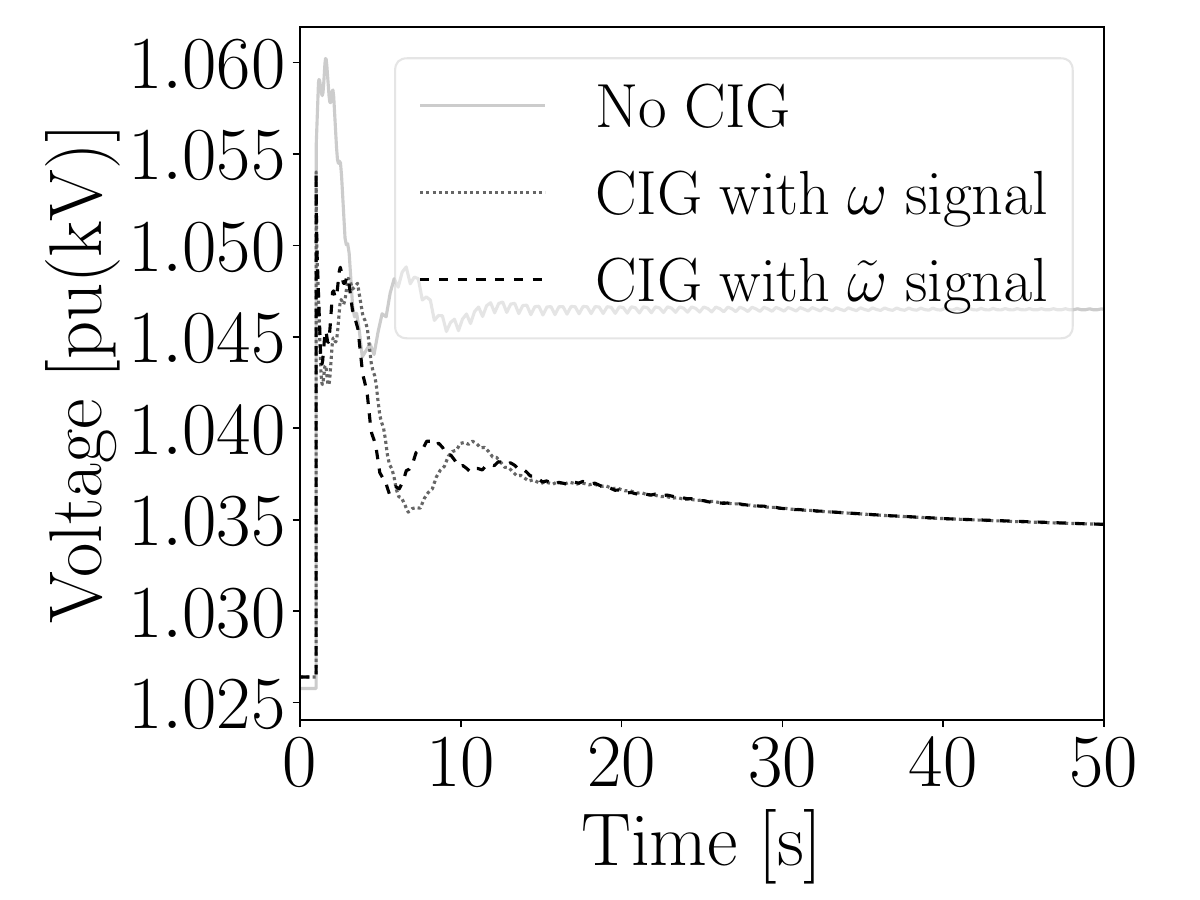}}
  \caption{Transient behavior of the frequency of the \ac{coi} and of
    the voltage at bus 7 for various control setups of the WSCC test
    system.}
  \label{fig:wv} 
  \vspace{-3mm}
\end{figure}

\begin{figure}[ht!]
  \centering
  \resizebox{0.49\linewidth}{!}{\includegraphics[scale=1.0]{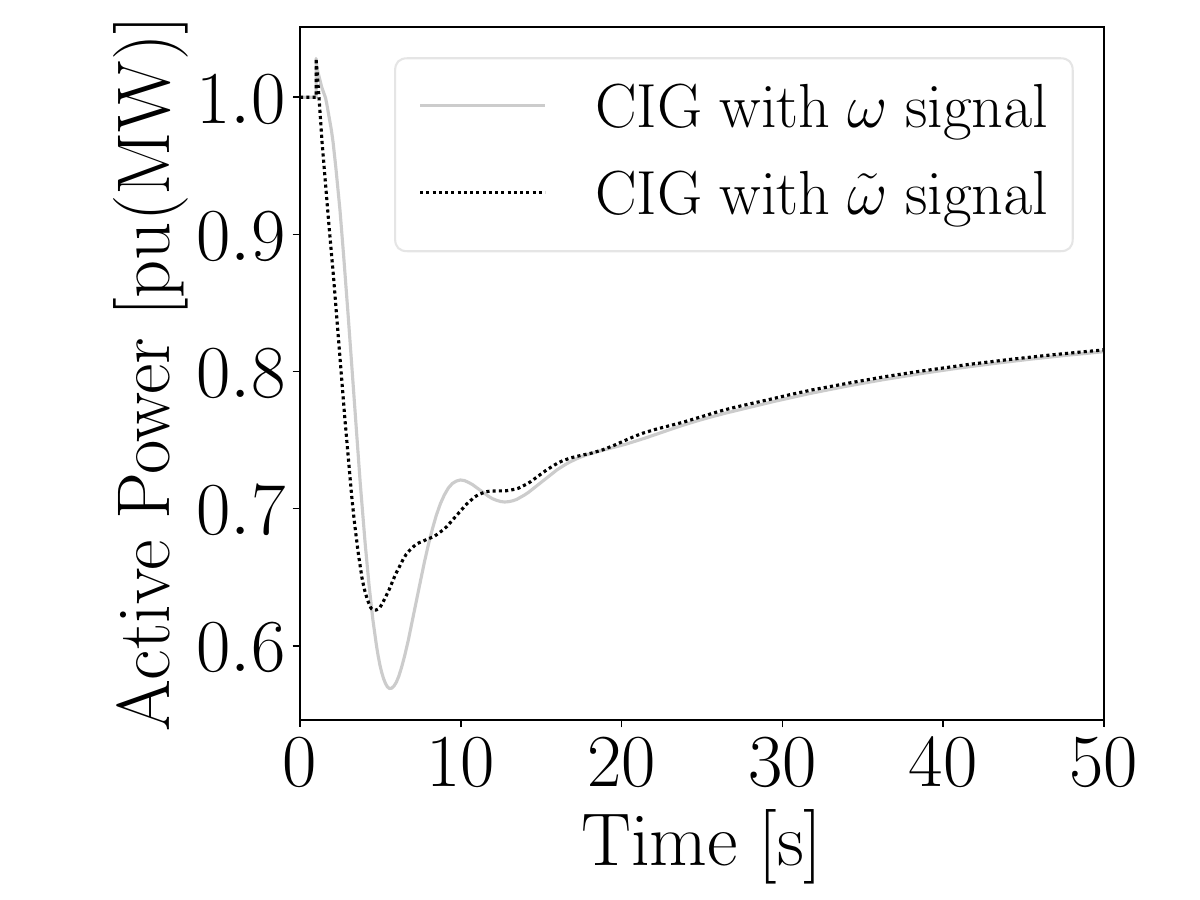}}
  \resizebox{0.49\linewidth}{!}{\includegraphics[scale=1.0]{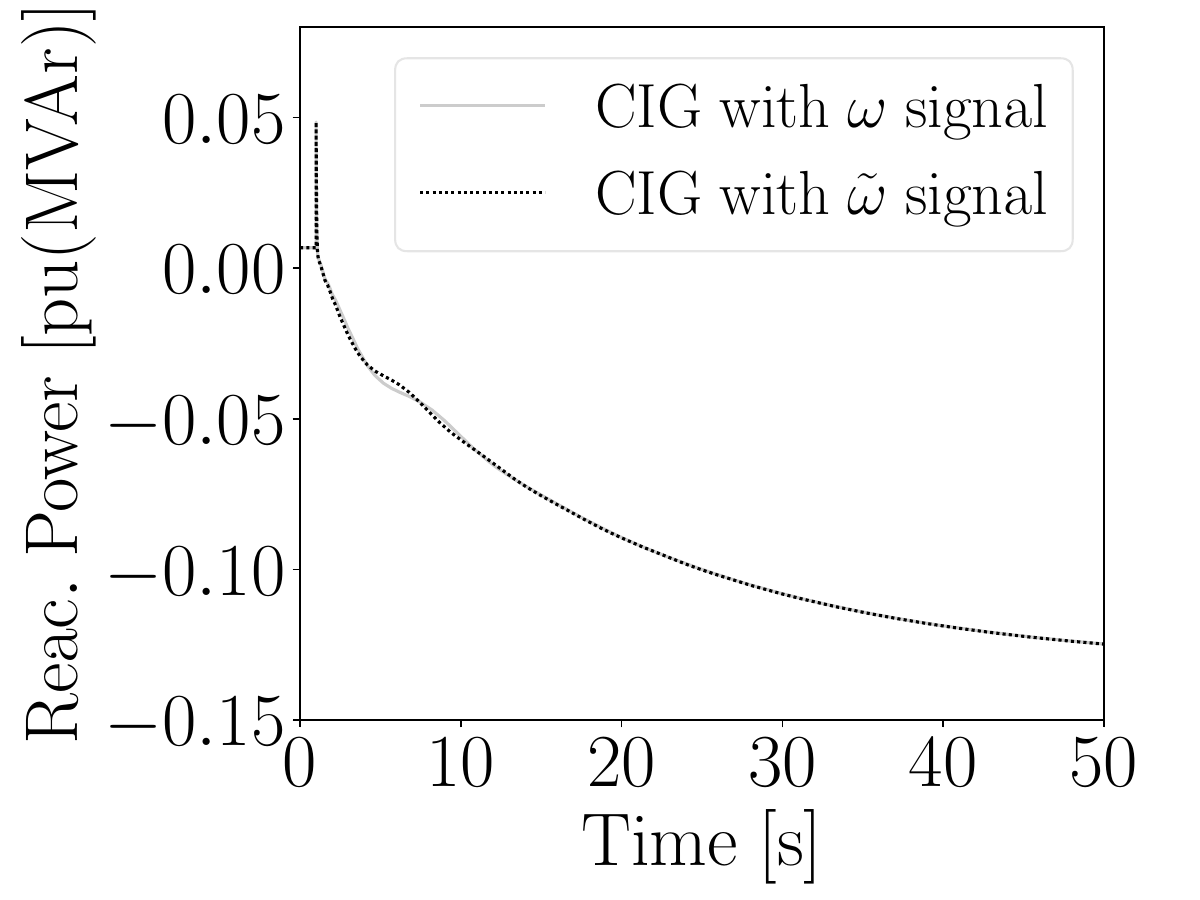}}
  \caption{Transient behavior of the power injected by the \ac{cig}
    for various control setups of the WSCC test system.}
  \label{fig:pq} 
  \vspace{-3mm}
\end{figure}

Figures \ref{fig:wv} and \ref{fig:pq} show the transient behavior of
the system following the loss of 50\% of the load consumption at
bus~5, for three scenarios: (i) system without \ac{cig}; (ii) system
with \ac{cig} and frequency control using the conventional signal
$\omega$; and (iii) system with \ac{cig} and frequency control using
the proposed signal $\tilde{\omega}$ defined in \eqref{eq:wt}.  The
results shown in Figs.~\ref{fig:wv} and \ref{fig:pq} were obtained for
$K=1.2$.

The proposed control achieves lower deviations of the \ac{coi}
frequency and of the power generated by the \ac{cig}, without
deteriorating the voltage control performance or modifying the
reactive power output.  The compensating signal has the sought effect
of reducing the local oscillation of the active power of the CIG,
which results in an overall improvement of the system frequency
dynamic response.

\subsection{New England 39-bus System}

\begin{figure}[ht!]
  \begin{center}
    \resizebox{0.75\linewidth}{!}{\includegraphics[scale=1.0]{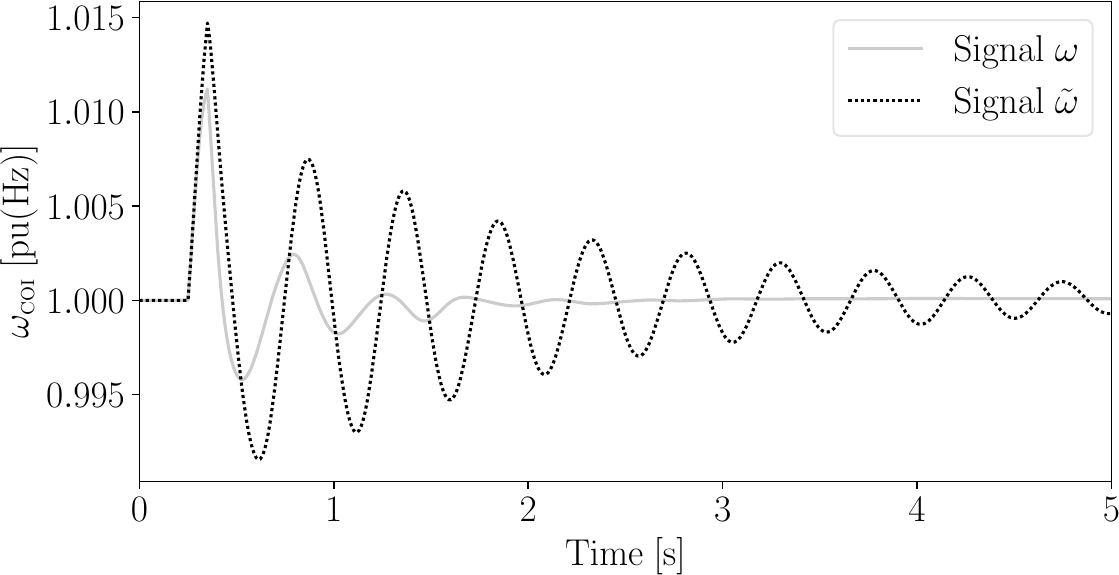}}
    \caption{New England 39-bus system: Transient behavior of the
      frequency for various setups of the frequency control of the
      \acp{cig}. }
    \label{fig:newengland}
  \end{center}
  \vspace{-4mm}
\end{figure}

This section illustrates the dynamic performance of the proposed
control for a larger system with multiple \acp{cig}.  The original
data is modified to accommodate 70\% of CIG-based generation and the
inertias of conventional \acp{sm} are reduced to emulate a low-inertia
system.  The setup of the grid is same as in
\cite{sanniti2022curvature}.  The contingency considered is a fault at
bus~12 cleared after $0.2$ s.  Figure \ref{fig:newengland} shows that,
also in this case, the compensating signal effectively reduces local
oscillations and improves the system frequency dynamic response.

It is relevant to note that, in this scenario, the signal
$\tilde{\omega}$ is obtained using $K=-0.03$.  Comparing this value
with that utilized for the WSCC 9-bus system ($K=1.2$), we note that
the gain $K$ is highly system-dependent and can be positive or
negative.  On the other hand, $K$ does not need to be adjusted when
the operating point changes and, according to our study, does not
appear to interfere or couple with other system dynamics.

% ======================================================================
\section{Conclusions}
\label{sec:Conclusion}
% ======================================================================

This letter builds on top of a recently proposed definition of
\textit{complex frequency} that accounts for angle and magnitude
voltage rate of changes in a unified geometric-based framework.  The
invariant properties of the components of the complex frequency are
exploited in this work to separate local and system-wide dynamics of
the frequency.  This separation allows defining a frequency control
that can compensate local oscillations and, as a consequence, improve
the overall transient response of the grid.  The effectiveness of this
approach is confirmed by the eigensensitivity analysis and the
evaluation of the geometric observability of the proposed compensated
control signal, and by time-domain simulations.  The proposed control
is suited for CIGs, the active power controllers of which are
generally faster than those of conventional power plants and are thus
more prone to be affected by local oscillations of the frequency.
Finally, the proposed control is also simple and inexpensive to
implement as it requires only local measurements of the voltage at the
point of connection of the devices that regulate the frequency.


% ======================================================================

\vfill

\end{document}